\documentstyle{neu}
\begin{document}

\title{{\it RXTE} Observations of the SGR 1806--20 Steady Emission}

\author{D. Marsden and R. E. Rothschild \\
{\it Center for Astrophysics and Space Sciences, University of 
California at San Diego, La Jolla, CA 92093} \\
C. Kouveliotou$^1$,S. Dieters$^2$ \\ 
{\it ES-84 NASA/MSFC \\ Huntsville, AL 35812} \\
J. van Paradijs$^2$ \\
{\it University of Amsterdam, Astronomical Institute 
``Anton Pannekoek'' \\ Kruislaan 403, 1098 SJ, Amsterdam} \\
$^1$Universities Space Research Association \\
$^2$University of Alabama, Huntsville}

\maketitle

\section{Introduction}
 
Soft gamma repeaters are astrophysical sources which exhibit long 
periods of quiescence, often spanning years, punctuated by periods of 
intense bursting activity during which many brief (durations $<1$ s) 
and intense (luminosities L$\sim1-10^{3}$ L$_{Edd}$) bursts are emitted 
by the source (Norris et al. 1991). Besides the burst emission, SGRs 
are also characterized by steady - also referred to as ``quiescent'' - 
emission. Believed to be neutron stars, the mechanism(s) for both the 
steady and bursting X--ray emission is still not well understood 
(Thompson \& Duncan 1995). 
 
SGR 1806-20 is the most prolific SGR, and it has been studied in the 
X--ray (Sonobe et al. 1994), optical (van Kerkwijk et al. 1995), 
infrared (Kulkarni et al. 1995), and radio (Kulkarni et al. 1994) bands. 
The source became active again during the Fall of 1996, emitting many 
powerful bursts that were first detected with BATSE (Kouveliotou et al. 
1996). A target of opportunity observation by the {\it Rossi X--ray 
Timing Explorer} (RXTE) was initiated on November 5, 1996. The data 
analyzed here were taken during that $50$ ks observation, which spanned 
the time interval starting at 10:53:20 UT (5/11/96) and ending at 
10:52:00 UT (6/11/96). In addition, followup {\it RXTE} data were 
taken during the time interval 06:45:20 UT (15/7/97) to 09:06:24 
(15/7/97), in which the source was not bursting.
 
\section{Spectral Analysis: {\it RXTE} data}

The pointed instruments aboard {\it RXTE} are the Proportional Counter 
Array (Jahoda et al. 1996) and the High Energy X--ray Timing Experiment 
(rothschild et al 1998). Only data from the PCA is used in this work, 
however, because the relatively faint quiescent emission from SGR 1806--20 
was not detected by the HEXTE on the short timescales discussed here. The 
PCA instrument consists of $5$ collimated Xenon proportional counter 
detectors with a total net area of $7000$ cm$^{2}$ and an effective energy 
range of $2-60$ keV. The instrumental background for the PCA is determined 
from modeling of both the internal background of the detectors and the 
background due to cosmic X-ray flux and charged particle events.

A $2-60$ keV lightcurve of the SGR 1806--20 {\it RXTE} TOO observation 
was extracted using standard FTOOLS 4.0 routines, and is shown in 
Figure 1. Spectra were extracted from two data stretches, of approximately 
equal duration, which contain no detectable burst emission from the SGR.
These regions, denoted ``A'' and ``B'', are marked by the dotted lines 
in Figure 1. In addition, a spectrum of approximately equal duration 
(column ``C'' in Table 1) was extracted from the July data, during 
which the source did not burst. Background spectra were generated for 
all three intervals using the standard background tool PCABACKEST. 

\begin{figure}[t]
\vspace{4.in}
\caption{The $2-60$ keV RXTE/PCA lightcurve of the SGR 1806--20 TOO 
observation. The vertical dotted lines denote the spectral accumulation 
intervals A and B. The background has not been subtracted, and the time
resolution is $0.5$ s.} 
\end{figure}

The PCA data from the {\it RXTE} observations were fit to specific 
functional forms over the energy range $2.5-20.0$ keV using XSPEC 10.00. 
The fitted shapes were of the form $N_{H}$ x continuum $+$ gaussian line, 
where $N_{H}$ is the neutral hydrogen photoelectric absorption column 
density (Morrison \& McCammon 1983). Raymond-Smith, thermal bremsstrahlung, 
and power law functional forms were used for the continuum spectral shape, 
but the power law form produced (by far) the best fit to the data. The 
spectral parameters for the Fe line in the three {\it RXTE} fits were 
consistent with a constant line normalization, centroid energy, and 
linewidth, with values of $(5.1\pm0.3) \times 10^{-4}$ 
Photons cm$^{-2}$ s$^{-1}$, $6.65\pm0.02$ keV, and $0.37\pm0.03$ keV, 
respectively, for these parameters. Assuming these constant Fe line 
parameters, which are due to emission from the galactic ridge (Yamauchi 
\& Koyama 1993), the best-fit {\it RXTE} spectral parameters for the 
three data intervals are given in Table 1. The spectral fit results 
indicate that the SGR 1806--20 steady emission is roughly consistent 
with a constant spectral shape and normalization throughout periods of 
both intense bursting and relative quiescence.

\begin{table}[t]
\caption{{\it RXTE} SGR 1806--20 quiescent spectral fit results} 
\vspace{.5pc}
\begin{center}
\begin{tabular}{lccc} \hline
\multicolumn{1}{c}{Parameter}& \multicolumn{1}{c}{A} & 
\multicolumn{1}{c}{B} & \multicolumn{1}{c}{C} \\
\hline
Flux$^{1}$ & $7.75\pm0.16$ & $7.83\pm0.18$ & 
$8.05\pm0.16$ \\
Photon Index & $2.29\pm0.03$ & $2.27\pm0.03$ & $2.33\pm0.03$ \\
N$_{H}$$^{2}$ & $2.9\pm0.3$ & $3.2\pm0.3$ & $3.1\pm0.3$ \\
${\chi_{\nu}}^{2}$ & $0.63$ & $0.61$ & $0.65$ \\
$\nu$ & $44$ & $44$ & $44$ \\
Livetime$^{3}$ & $3300$ & $2600$ & $3600$ \\ \hline
\end{tabular}
\end{center}
\hspace{1.in}$^{1} 2-10$ keV power law flux ($10^{-11}$ ergs cm$^{-2}$ 
s$^{-1}$)

\vspace{-.15pc}
\hspace{1.in}$^{2}$ Neutral hydrogen absorption ($10^{22}$ H Atoms 
cm$^{-2}$)

\vspace{-.10pc}
\hspace{1.in}$^{3}$ Instrumental livetime for spectral fit (seconds) \\  
\end{table}
 
\section{Conclusion}

Through analysis of {\it RXTE} TOO data, we have determined that the 
persistent emission from SGR 1806--20 is consistent with a constant 
spectral shape and intensity both during and away from the active 
bursting periods. The spectrum is best-fit by a nonthermal power 
law shape, with thermal bremsstrahlung and Raymond-Smith functional 
forms producing much worse fits to the {\it RXTE} data. The mean 
power law spectral index obtained by {\it RXTE} is $2.30\pm0.02$, 
which is consistent with the {\it ASCA} value of $\alpha=2.2\pm0.2$ 
(Sonobe et al. 1994).

The nonthermal nature of the SGR 1806--20 quiescent spectrum supports 
the idea that the X--ray emission is due to a compact synchrotron 
nebula, or plerion, that derives its power from either a rapidly 
spinning-down pulsar (Kulkarni et al. 1994) or energetic particles 
ejected in the SGR bursts (Tavani 1994). The lack of an iron line in 
the SGR spectrum (Sonobe 1994) and short term time variability argue 
against the source being a low luminosity X--ray binary system (White, 
Nagase, \& Parmar 1995), although a more extensive monitoring campaign 
is necessary to rule out this hypothesis. 
 
\paragraph*{Acknowledgements.} We thank NASA for support under grants
NAS5-30720 (D.M. and R.E.R.), NAG5-2560 (S.D. and C.K.), and NAG5-4878 
(JvP)

\section*{References}

\vspace{1pc}

\re
Jahoda, K. et al. 1996, EUV, X--ray, and Gamma--Ray Instrumentation 
for Astronomy VII, SPIE Proceedings, eds: O. H. V. Sigmund and M. Gumm, 
2808, 59
\re
Kouveliotou, C. et al 1996, {\it IAUC}, 6501
\re
Kulkarni, S. R. et al. 1994, {\it Nature}, 368, 129
\re
Kulkarni, S. R. et al. 1995, {\it ApJ}, 440, L61
\re
Morrison, R. and McCammon, D. 1983, {\it ApJ}, 270, 119
\re
Norris, J. P. et al. 1991, {\it ApJ}, 366, 240
\re
Rothschild, R. E. et al. 1998, {\it ApJ}, 496, in press
\re
Sonobe, T. et al. 1994, {\it ApJ}, 436, L23
\re
Tavani, M. 1994, {\it ApJ}, 431, L83
\re
Thompson, C. \& Duncan, R. C. 1995, {\it MNRAS}, 275, 255
\re
van Kerkwijk, M. H.  et al. 1995, {\it ApJ}, 444, L33
\re
White, N. E., Nagase, F., and Parmar, A. N. 1995 
in X--ray Binaries, eds. W. H. G. Lewin, J. van Paradijs, and 
E. P. J. van den Heuvel (Cambridge University Press: Cambridge), 1
\re
Yamauchi, S. and Koyama, K. 1993, {\it ApJ}, 404, 620

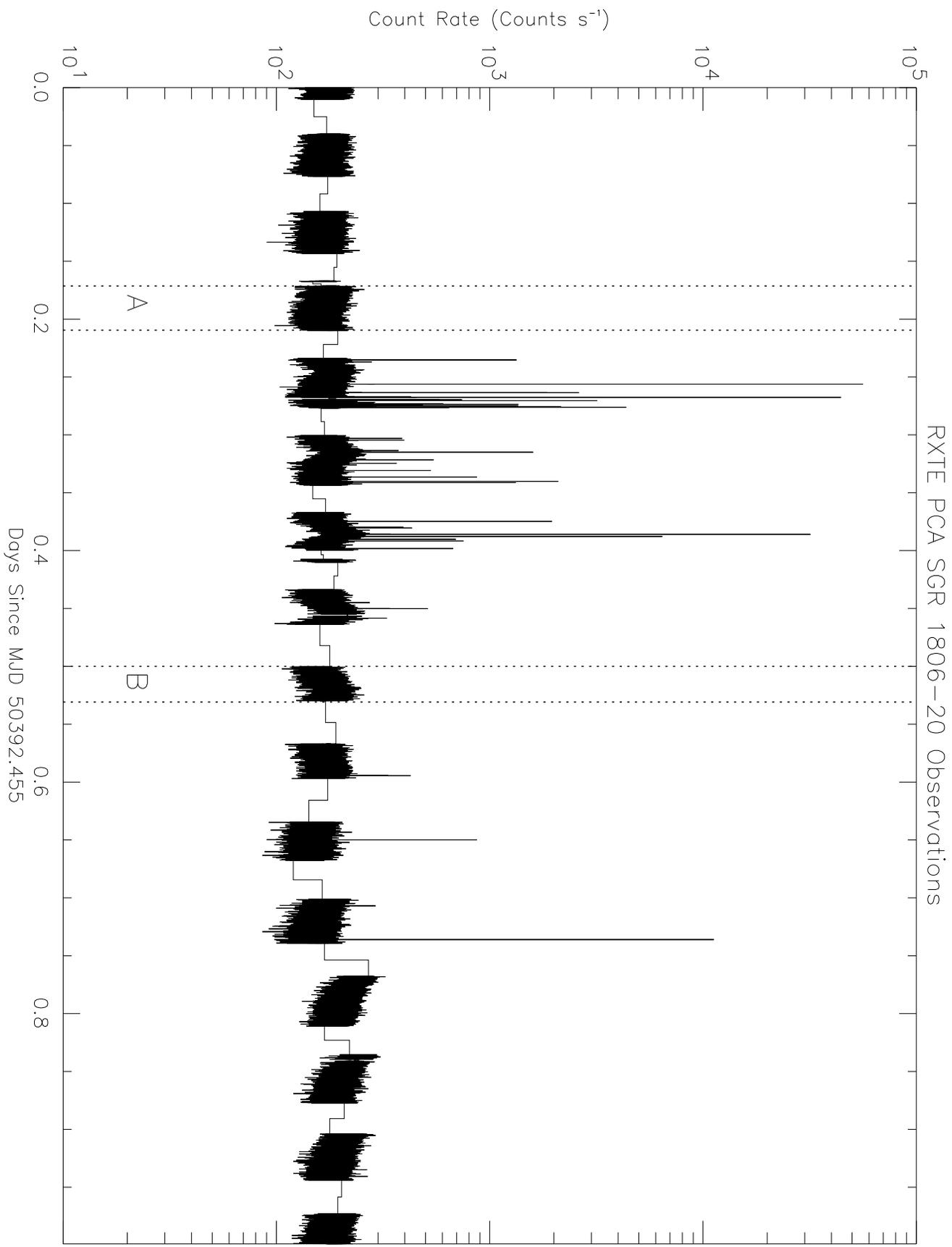
\end{document}